\documentclass[conference]{IEEEtran}
\IEEEoverridecommandlockouts

\usepackage{cite}
\usepackage{amsmath,amssymb,amsfonts}
\usepackage{algorithmic}
\usepackage{graphicx}
\usepackage{textcomp}
\usepackage{xcolor}
\usepackage{epsfig,epstopdf,graphics,subfigure,psfrag,amsmath,cases}
\usepackage{latexsym,amssymb,amsmath,epsfig,subfigure,algorithm,amsthm}
\usepackage{algorithmic}
\usepackage{color}
\usepackage{url}
\usepackage{scrtime}
\usepackage{bm}
\usepackage{graphicx}
\usepackage{hyperref}
\usepackage{subfigure}
\usepackage{bbding}
\usepackage{multicol}
\usepackage{graphics}
\usepackage{amsmath}
\usepackage{amsthm}
\usepackage{cite}
\usepackage{array}
\usepackage{verbatim}

\def\BibTeX{{ B\kern-.05em{\sc i\kern-.025em b}\kern-.08em
		T\kern-.1667em\lower.7ex\hbox{E}\kern-.125emX}}

\begin{document}
	
	\title{ Beamforming Design for Secure MC-NOMA Empowered ISAC Systems with an Active Eve 
    \thanks{This work was supported by the Beijing Natural Science Foundation
			under Grant L222004 and the Young Backbone Teacher Support Plan of Beijing Information Science $\&$ Technology University  under Grant YBT 202419.}
	}
	
\begin{comment}

\end{comment}

\author{\IEEEauthorblockN{Zhongqing Wu$^*$, Xuehua Li$^*$,  Yuanxin Cai$^*$, Weijie Yuan\dag}
	\IEEEauthorblockA{*{Key Laboratory of Information and Communication Systems, Ministry of Information Industry,} \\
	{Beijing Information Science and Technology University, 	Beijing, China}\\
	}
		\IEEEauthorblockA{\dag{Department of Electrical and Electronic Engineering,
			} \\
		{	Southern University of Science and Technology, Shenzhen, China}\\
		Email:  \{zhongqing.wu, lixuehua,  cai$\_$yuanxin\}@bistu.edu.cn, yuanwj@sustech.edu.cn}
	}

	\maketitle

\begin{abstract}
As the integrated sensing and communication (ISAC) technology emerges as a promising component of  sixth-generation (6G),
the study of its physical layer security has become a  key concern for researchers.
Specifically, in this work, we focus on the security issues over a multi-carrier (MC)-non-orthogonal multiple access (NOMA) assisted ISAC system, considering imperfect channel state information (CSI) of an active Eve and graded confidentiality demands for users.
To this end, the subcarrier allocation, the information, and artificial noise beamforming are designed to maximize  the minimum   communication rate, while ensuring diverse confidentiality and sensing performance demands.
An effective security strategy is devised via the Lagrangian dual transformation and successive convex approximation methods.
Simulations confirm  the validity and robustness   of the proposed scheme in terms of the security performance.

\end{abstract}


\section{Introduction}
Due to the enhanced spectrum utilization, reduced hardware cost and signalling overhead of the integrated sensing and communication (ISAC), it has been regarded as a foreseeable technique for sixth-generation (6G) mobile networks\cite{wei2022toward,liu2022survey,liao2021design}.
Nevertheless, the shared nature of ISAC also makes the confidential signals more susceptible to untrusted targets (also known as Eves)\cite{su2023security}.
Recently, reliable ISAC has received growing research interest.
For instance, the authors in \cite{chu2022joint} proposed a joint secure transmit beamforming design to minimize the maximum eavesdropping SINR and ensure reliable communication in ISAC systems.
Moreover, in \cite{xu2022robust}, the artificial noise (AN) and information beamforming vector are jointly designed to maximize the sum secrecy rate (SSR) based on variable-length snapshots.
However, previous works mostly considered the passive eavesdropping threat from the Eve  \cite{chu2022joint,xu2022robust,bazzi2023secure,chu2023joint}, while overlooking the malicious jamming attacks from an active Eve, thereby compromising the reliability and security.

Meanwhile, non-orthogonal multiple access (NOMA) has been integrated into the ISAC systems to accommodate more users and enhance resource utilization.
To guarantee security for NOMA users, the authors of \cite{luo2022secure} performed legitimate users’ throughput maximization design via cooperative radar jamming.
However, user fairness over the data rate is neglected in this work.
In \cite{huang2024secure}, the SSR of users is maximized by providing secure precoding under perfect and imperfect channel state information (CSI) of  Eve. 
Yet, the above studies only consider a single subcarrier, which fails to fully exploit the benefits of NOMA in ISAC systems.
Furthermore, despite the effective security solutions in existing research, there is still a notable lack of studies on graded security protection, which is an urgent need in practical multi-user ISAC applications.

In this paper, a multi-carrier (MC)-NOMA empowered ISAC system is investigated to maximize the minimum communication rate, different from\cite{luo2022secure}, which ignores the user-fairness, ensuring users'  graded security needs and basic sensing performance.
We propose an SCA-based iterative algorithm to jointly optimize the secure beamforming and subcarrier scheduling.
The effectiveness and superiority of the algorithm are validated in the simulation results.

	\section{System Model}
\begin{figure}[htbp] \vspace*{-3mm}
	\centering
	\includegraphics[width=2.4 in]{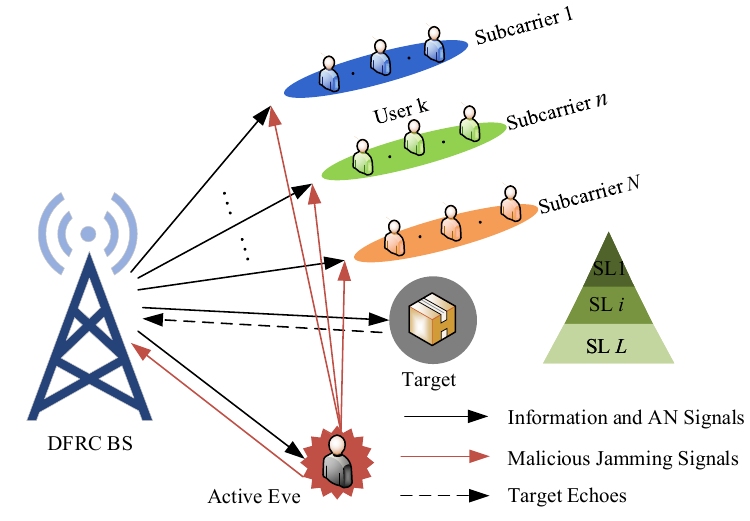} 
	\caption{The considered MC-NOMA empowered ISAC system.}
	\label{system_model} \vspace*{-2mm}
\end{figure}
We consider an MC-NOMA-enabled ISAC system as shown in Fig. \ref{system_model}, where a dual-functional radar communications (DFRC) base station (BS), equipped with uniform linear arrays (ULAs) of $V$ antennas, serves $K$ single-antenna users with the help of NOMA, denoted by $ {{\cal K}} = \{1,...,k,...,K\} $. We consider a point-like target of interest and  an active Eve  in the system.
We assume that the active Eve is equipped with $M$ ULA antennas. 
The malicious jamming signals from the active Eve can be represented as
$\textbf{x}_{E}=\sqrt {{P_{E}}} {\textbf{x}_{se}}\in {\mathbb{C}^{M \times 1}},$
where $P_{E}$ and  ${{\textbf{x}}_{se}} \sim\mathcal{CN}(\textbf{0},\textbf{I}_M)$ denote the maximum transmit power and launched signals at the active Eve, respectively.
Note that we consider  $L$   security levels (SLs) for communication users in this work, where different confidentiality requirements are predefined for specific applications.
Besides, a wideband system is considered in this paper, which is divided into $N$ orthogonal subcarriers with each bandwidth $B$.
We introduce a user scheduling variable $s_{n,k}\in \{0,1\}, n\in {{\cal N}}  \buildrel \Delta \over =\{1,...,n,...,N\}$, $k\in {{\cal K}}$ to denote that user $k$ is grouped to the subcarrier $n$ when $s_{n,k}=1$ and $s_{n,k}=0$ otherwise.
Besides, we assume that at most $K_N$ users are allocated to each subcarrier as a cluster. 
Note that BS knows the perfect CSI of all communication users  and the approximate sensing range of the target while considering imperfect CSI for the  Eve, which is widely adopted in related works \cite{hu2022beamforming,hu2023secure}.

\subsection{Signal Model}
To guarantee secure communication, the AN is superimposed into the information-bearing DFRC signals.
Therefore, the signals transmitted from BS on  subcarrier $n$ is given by
\begin{align}
	{\textbf{x}_n} = \sum\limits_{k = 1}^{K} {{s_{n,k}}} {\textbf{w}_{n,k}}{c_{n,k}}+\textbf{v}_nz,
\end{align}
where $\textbf{w}_{n,k}\in {\mathbb{C}^{V \times 1}}$ denotes the  beamforming vector for the  user $k$ on  subcarrier $n$ and ${c_{n,k}}$ is the corresponding information symbol, with ${\left| {{c_{n,k}}} \right|^2} = 1$, $ \mathbb{E}\{ c_{n,j}c_{n,k}^T\}  = 0, j \ne k$.
Note that $\textbf{v}_n\in {\mathbb{C}^{V \times 1}}$ denotes the AN beamforming vector on subcarrier $n$ and $z \sim\mathcal{CN}(0,1)$ is the random AN symbol with unit power ${\left| z \right|^2} = 1$ and $\mathbb{E}\{ c_{n,k}z^T\}  = 0$.

\subsection{Communication Model}
The received signals at user $k$ and the active Eve on subcarrier $n$ can be expressed as
\begin{align}\label{signal}
	{y_{n,k}} &= {\textbf{h}^{\mathrm H}_{n,k}}\textbf{x}_n +(\textbf{h}_{n,k}^{E})^{\mathrm H}{\textbf{x}_{E}} + {n_{n,k}},\,\,\text{and}\\
	\label{eve_radar}
	{\textbf{y}_{n}^{E}} &=  {({\textbf{H}}_{n}^{BE})^{\mathrm{H}}{\textbf{x}_n}}  +\textbf{n}_{E}, 
\end{align}
respectively, where  ${\textbf{h}_{n,k}}\in {\mathbb{C}^{V \times 1}}$   denotes the channel  from the BS  to user $k$ on subcarrier $n$, which is modeled as Rayleigh fading in this work.
In addition, we assume that the rough  channel estimation of the active Eve, i.e., $\widehat{\textbf{H}}_{n}^{BE}$ and $\widehat{\textbf{{h}}}^{E}_{n,k}$,   can be obtained via the application of cognitive paradigms as in \cite{9133130}.
In light of this, the channels from BS to the active Eve $\textbf{H}_{n}^{BE}\in {\mathbb{C}^{V \times M}}$, as well as from the active Eve to user $k$, ${\textbf{{h}}}^{E}_{n,k}\in {\mathbb{C}^{M \times 1}}$     can be modelled by
\begin{align}
{\textbf{H}}_{n}^{BE}&=\widehat{\textbf{H}}_{n}^{BE}+\Delta \textbf{H}_{n}^{BE},\\
{\textbf{{h}}}^{E}_{n,k}&=\widehat{\textbf{{h}}}^{E}_{n,k}+\Delta {\textbf{{h}}}^{E}_{n,k},
\end{align}
where $\bm  \Omega_{n}^{BE}  \buildrel \Delta \over = \{ \Delta {{\textbf{H}}_{n}^{BE}} \in {\mathbb{C}^{V \times M}}:\left\| {\Delta {{\textbf{H}}_{n}^{BE}}} \right\|_F \le {\varepsilon _{n}^{BE}}\} \,\,\text{and}\,\,
\bm \Omega^{E}_{n,k}  \buildrel \Delta \over = \{ \Delta {{\textbf{h}}}^{E}_{n,k} \in {\mathbb{C}^{M \times 1}}:{\left\| {\Delta {\textbf{h}}_{n,k}^{E}} \right\|_2} \le {\varepsilon ^{E}_{n,k}}\}
$ denote  all the possible channel estimation errors in  the continuous sets, respectively,
where the constants ${\varepsilon _{n}^{BE}} \,\,\text{and}\,\,{\varepsilon ^{E}_{n,k}}$ denote the maximum CSI estimation error value for the corresponding channel, respectively.
Besides, $ {n_{n,k}}\sim\mathcal{CN}(0,\sigma _{n,k}^2)$ and $ {\textbf{n}_{E}}\sim\mathcal{CN}(\textbf{0},\sigma _{E}^2\textbf{I}_M)$ represent the additive white Gaussian noise (AWGN) of $k$ and the active Eve, respectively.

Without loss of generality, we assume that the successive interference cancellation (SIC) process can be conducted  successfully as user $k$ received the signal $y_{n,k}$.
Specifically, the user $k$ first decodes and removes the signals from the weaker users,  and then  regards the multi-user interference (MUI) from the stronger users as noise.
Accordingly, we yield the signal-to-interference-plus-noise ratio (SINR)  for the user $k$  on subcarrier $n$\cite{9682500} 
\begin{align}\label{SINR_K}
\text{SINR}_{n,k} \!\!=\!\! \frac{{{{s_{n,k}}{\left| {{\textbf{h}}_{n,k}^{\mathrm H}{{\textbf{w}}_{n,k}}} \right|^2}}}}{{{\text{MUI}}_{n,k} + {\left| {{{(\textbf{h}_{n,k}^{E})}^\mathrm{H}}{\textbf{x}_{E}}} \right|^2} +{\left|{\textbf{h}_{n,k}^{\mathrm H}\textbf{v}_n}\right|^2} + \sigma _{n,k}^2}},
\\[-7.5mm]\notag
\end{align}
where ${\text{MUI}}_{n,k}=\sum\limits_{{\left\| {{\textbf{h}_{n,j}}} \right\|^2} > {\left\| {{\textbf{h}_{n,k}}} \right\|^2}} {s_{n,j}}{\left| {{\textbf{h}}_{n,k}^{\mathrm H}{{\textbf{w}}_{n,j}}} \right|^2}$.
Thus, the achievable rate for user $k$ can be given by
\begin{align}
{R}_{k} = \sum\limits_{n = 1}^N {\log _2}(1 + \text{SINR}_{n,k}),\forall k.
\end{align}

Without loss of generality,  we assume that the active Eve is capable of eliminating the multi-user interference via perfect SIC,  drawing the theoretical lower bound of the secrecy performance for our considered system.
Thus, the leakage SINR for user $k$ on subcarrier $n$ can be expressed as
\begin{align}\label{SINR_{E}}
\text{SINR}_{n,k}^{E} = \frac{{{s_{n,k}}{{\left\| {({\textbf{H}}_{n}^{BE})^\mathrm{H} {{\textbf{w}}_{n,k}}} \right\|}^2} }}{{{{\left\| {({\textbf{H}}_{n}^{BE})^\mathrm{H} {\textbf{v}_n}} \right\|}^2}  + \sigma _{E}^2}},
\end{align}
thus yielding the corresponding leakage rate of the user $k$
\begin{align}
R_k^{E} = \sum\limits_{n = 1}^N  {\log _2}(1 + \text{SINR}_{n,k}^{E}). 
\end{align}

\subsection{Sensing Model }
To begin with, the received echos expression at the BS  on subcarrier $n$, $\textbf{y}_{n}^{R}\in {\mathbb{C}^{V \times 1}}$,  can be derived by
\begin{align}\label{BS_radar}
{{\textbf{y}}_{n}^{R}} &= {\beta _t}{\textbf{H}_{n}^{BT}}{{\textbf{x}}_n}  + \textbf{H}_{n}^{BE}{{\textbf{x}}_{E}} + {{\textbf{n}}_r},
\end{align}
where ${\textbf{H}_{n}^{BT}} = {\textbf{h}_{n}^{BT}}(\textbf{h}_{n}^{BT})^\mathrm{H}\in {\mathbb{C}^{V \times V}}$, and  ${\textbf{h}_{n}^{BT}}\in {\mathbb{C}^{V \times 1}}$  denotes the channel  from BS  to  the  target on subcarrier $n$, which is modeled as Rayleigh fading.
Note that $\beta_t$ denotes the reflection coefficient based on the target cross-section. 
Besides, ${\textbf{n}_{r}}\sim\mathcal{CN}(0,\sigma _{r}^2\textbf{I}_V)$  represents the corresponding AWGN, with the noise power  $\sigma _{r}^2$.
Therefore, we yield the  interference-plus-noise covariance matrix on subcarrier $n$ as following
\begin{align}
 { \bm {\mathcal{R}}}_\mathrm{n} &= {\textbf{H}_{n}^{BE}}\mathbb{E}\{ {{\textbf{x}}_{E}}{{\textbf{x}}^\mathrm{H}_{E}}\} ({\textbf{H}}_{n}^{BE})^\mathrm{H} + \sigma _r^2{{\textbf{I}}_V}\in {\mathbb{C}^{V \times V}}\\ \notag
& =  {P_{E}}{{\textbf{H}}_{n}^{BE}}({\textbf{H}}_{n}^{BE})^\mathrm{H} + \sigma _r^2{{\textbf{I}}_V}.
\end{align}

In terms of the target sensing, one of the main tasks is  to estimate the direction-of-arrival (DOA) for target, i.e., ${\theta _t}$. 
Moreover, Cram\'er-Rao bound (CRB) is adopted to evaluate our parameter estimation performance, which characterizes the lower bound for the mean square error.
Similar to \cite{10086570}, 
the CRB related to $ {\theta _t}$ for the BS can be formulated as
\begin{align}\notag
\text{CRB}({\theta _t})&\!=
\sum\limits_{n = 1}^N \!\! {\frac{1}{{2\Re \{ {\text{Tr}}[\frac{{\partial {{({\beta _t}{{\textbf{H}}_{n}^{BT}}{{\textbf{x}}_n})}^{\mathrm{H}}}}}{{\partial {\theta _t}}}{ \bm {\mathcal{R}}}_\mathrm{n}^{ - 1}\frac{{\partial ({\beta _t}{{\textbf{H}}_{n}^{BT}}{{\textbf{x}}_n})}}{{\partial {\theta _t}}}]\} }}} \\
&= \sum\limits_{n = 1}^N {\frac{1}{{2{{\left| {{\beta _t}} \right|}^2}}}} {({\text{Tr}}[{\textbf{x}}_n^{\mathrm{H}}({{\dot {\textbf H}}}_{n}^{BT})^\mathrm{H}{ \bm {\mathcal{R}}}_\mathrm{n}^{ - 1}{{{{\dot {\textbf H}}}}_{n}^{BT}}{{\textbf{x}}_n}])^{ - 1}},
\end{align}
where ${{{{\dot {\textbf H}}}}_{n}^{BT}} = {{{{\dot {\textbf h}}}}_{n}^{BT}}({\textbf{h}}_{n}^{BT})^{\mathrm{H}} + {{\textbf{h}}_{n}^{BT}}({{\dot {\textbf h}}}_{n}^{BT})^{\mathrm{H}}$ and ${{{{\dot {\textbf h}}}}_{n}^{BT}} = \frac{{\partial {{\textbf{h}}_{n}^{BT}}}}{{\partial {\theta _t}}}$.

\section{Problem Formulation}
We aim to jointly design the subcarrier scheduling  and the secure beamforming to
maximize  the minimum  communication rate, while guaranteeing  the graded  security  sensing needs under imperfect CSI of the active Eve.
Therefore, the joint subcarrier allocation and beamforming design for the  considered system can be formulated as 
\begin{align} & \label{proposed_formulation_origion} \underset{s_{n,k},{\textbf{w}}_{n,k},\textbf{v}_n}{\mathrm{maximize}}\,\, 
\mathop {\min\limits_{k \in \cal K} {R_k}} 
\\[-1.5mm]
\mathrm{s.t.} \, \mathrm{C1}&\!: {s_{n,k}} \in \{ 0,1\} ,\forall n,k,	\notag \hspace{4mm} 
\mathrm{C2}\!: 	2 \le \sum\limits_{k = 1}^K {{s_{n,k}} \le {K_N}} ,\forall n, 
 \\[-1.5mm]\notag
\mathrm{C3}&\!:0 \le \sum\limits_{n = 1}^N ({\sum\limits_{k = 1}^K {{s_{n,k}}} {{\left\| {{{\textbf{w}}_{n,k}}} \right\|}^2}}  + {\left\| \textbf{v}_n \right\|^2} )\le {P^{\text{max}}_{BS}},  \\[-1.5mm]\notag
\mathrm{C4}&\!: \mathop {\min }\limits_{\bm \Omega^{E}_{n,k} } {R_k}  \ge {R_{\text{min} }},\forall k,
 \\[-1.5mm]\notag
\mathrm{C5}&\!: \mathop {\max }\limits_{\bm  \Omega_{n}^{BE}} R_k^{E} \le R_{{\text{leakage}}}^i,\forall k,\forall i \in \{ 1,...,L\},  \\[-1.5mm]\notag
\mathrm{C6}&\!:	\mathop {\max }\limits_{\bm  \Omega_{n}^{BE}} {\text{CRB}}({\theta _t}) \le {\Gamma _{\text{CRB}}},
\end{align}
where the binary constraint for the subcarrier scheduling variable is given by $\mathrm{C1}$.
$\mathrm{C2}$ denotes that  at most $K_N$ users are assigned to the same subcarrier. 
Besides, the maximum transmit power of the BS is denoted by $ {P^{\text{max}}_{BS}}$ in  $\mathrm{C3}$.
$\mathrm{C4}$ guarantees the fundamental communication performance for each user under imperfect CSI of the active Eve.
More remarkably,  different threshold values of the maximum tolerable leakage rates are defined in $\mathrm{C5}$, to cater for various confidentiality needs.
Moreover, the maximum tolerated distortion at the BS,  ${\Gamma _{\text{CRB}}} $,  is defined in $\mathrm{C6}$  to ensure the parameter estimation accuracy of the target.

\section{Problem Solution}
To facilitate our design, 
we first  define ${\textbf{W}_{n,k}} = {\textbf{w}_{n,k}}\textbf{w}_{n,k}^{\mathrm H} \in \mathbb{H}{^V},  \textbf{V}_n = \textbf{v}_n{\textbf{v}_n^{\mathrm H}}\in \mathbb{H}{^V},$ and $
{\textbf{H}_{n,k}} = {\textbf{h}_{n,k}}\textbf{h}_{n,k}^{\mathrm H}\in \mathbb{H}{^V}$, so that the SINR expressions in  \eqref{SINR_K} and \eqref{SINR_{E}} can be recast as 
\begin{align}
{\text{SINR}_{n,k}} &= \frac{{s_{n,k}}\text{Tr}({\textbf{H}_{n,k}}{\textbf{W}_{n,k}})}{{{\text{MUI}}_{n,k} + {\left| {{{(\textbf{h}_{n,k}^{E})}^\mathrm{H}}{\textbf{x}_{E}}} \right|^2} + \text{Tr}({\textbf{H}_{n,k}}{\textbf{V}_n}) + \sigma _{n,k}^2}},\text{and}\\
{\text{SINR}}_{n,k}^{E} &= \frac{{{s_{n,k}}{\text{Tr}}({\textbf{H}}_{n}^{BE}({{\textbf{H}}_{n}^{BE}})^{\mathrm{H}}{{\textbf{W}}_{n,k}}) }}{{{\text{Tr}}({\textbf{H}}_{n}^{BE}({{\textbf{H}}_{n}^{BE}})^{\mathrm{H}}{\textbf{V}_n})  + \sigma _{E}^2}}, \forall n,k,\,\,\text{respectively},
\end{align} 
where ${\text{MUI}}_{n,k}=\sum\limits_{{\left\| {{\textbf{h}_{n,j}}} \right\|^2} > {\left\| {{\textbf{h}_{n,k}}} \right\|^2}} {s_{n,j}}\text{Tr}({\textbf{H}_{n,k}}{\textbf{W}_{n,j}}) $.

In the sequel, note that the variables $s_{n,k}$ and $\textbf{W}_{n,k}$ are tightly coupled.
To tackle this issue, we introduce an auxiliary variable $\overline {\textbf{W}}_{n,k} = {s_{n,k}}{\textbf{W}_{n,k}}$ to decouple these  variables via adopting the Big-M formulation\cite{cai2022resource}
\begin{align}
\mathrm{C7a}&:{\overline {\textbf{W}} _{n,k}} \preceq {s_{n,k}}P_{BS}^{\max }{\textbf{I}_V},\forall n,k,\\
\mathrm{C7b}&:{\overline {\textbf{W}} _{n,k}} \succeq {\textbf{W}_{n,k}} - (1 - {s_{n,k}})P_{BS}^{\max }{\textbf{I}_V},\forall n,k,\\
\mathrm{C7c}&:{\overline {\textbf{W}} _{n,k}} \preceq {\textbf{W}_{n,k}},\forall n,k,
\mathrm{C7d}:{\overline {\textbf{W}} _{n,k}} \succeq \textbf{0},\forall n,k,
\end{align}
Then, through several mathematical transformations,
the constraints $\mathrm{C4}$ and $\mathrm{C5}$ can be equivalently
converted into the following manageable forms 
\begin{align}
\overline{\mathrm{C4}}&:\mathop {\min }\limits_{\bm \Omega^{E}_{n,k} }{\overline{R}_k}=
\mathop {\min }\limits_{\bm \Omega^{E}_{n,k} } \sum\limits_{n = 1}^N  {{{\log }_2}(1 + \overline{\text{SINR}}_{n,k}) \ge {R_{{\text{min}}}},\forall k,} \,\,\text{and}\\
\overline{\mathrm{C5}}&: \mathop {\max }\limits_{\bm  \Omega_{n}^{BE}} \sum\limits_{n = 1}^N {{{\log }_2}(1 +  \overline{\text{SINR}}_{n,k}^{E}) \le R_{{\text{leakage}}}^i,} \forall k,\forall i,  
\end{align}
respectively, where 
\begin{align}
\overline{\text{SINR}}_{n,k} &=\frac{\text{Tr}({\textbf{H}_{n,k}}{\overline {\textbf{W}} _{n,k}})}{{{\text{MUI}}_{n,k} +  \left| {{{(\textbf{h}_{n,k}^{E})}^\mathrm{H}}{\textbf{x}_{E}}} \right|^2 + \text{Tr}({\textbf{H}_{n,k}}{\textbf{V}_n}) + \sigma _{n,k}^2}},\\
\overline{\text{SINR}}_{n,k}^{E} &=\frac{{{\text{Tr}}({\textbf{H}}_{n}^{BE}{({\textbf{H}}_{n}^{BE})^\mathrm{H}}{{\overline {\textbf{W}} }_{n,k}}) }}{{{\text{Tr}}({\textbf{H}}_{n}^{BE}{({\textbf{H}}_{n}^{BE})^\mathrm{H}}{\textbf{V}_n}) + \sigma _{E}^2}}= \frac{{\xi_{n,k}^n}}{{\xi _{n,k}^d}},
\end{align}
where ${\text{MUI}}_{n,k}=\sum\limits_{{\left\| {{\textbf{h}_{n,j}}} \right\|^2} > {\left\| {{\textbf{h}_{n,k}}} \right\|^2}} {{\text{Tr}}({{\textbf{H}}_{n,k}}{{\overline {\textbf{W}} }_{n,j}})} $.

Therefore, the problem in \eqref{proposed_formulation_origion} can be reformulated as
\begin{align}& 
\label{recast_problem}
\underset{s_{n,k},{\textbf{W}}_{n,k},{{\overline {\textbf{W}} }_{n,k}},\textbf{V}_n,\textbf{z}_n}{\mathrm{maximize}}\,\, 
\mathop {\min\limits_{k \in \cal K} {\overline{R}_k}} \\\notag
\mathrm{s.t.} \, &\mathrm{C1},\mathrm{C2}, \overline{\mathrm{C4}},  \overline{\mathrm{C5}},
\mathrm{C7a-C7d},\\\notag
\overline{\mathrm{C3}}&\!:
0 \le \sum\limits_{n = 1}^N( {\sum\limits_{k = 1}^K {\text{Tr}({{\overline {\textbf{W}} }_{n,k}})} }  + \text{Tr}(\textbf{V}_n)) \le P_{BS}^{{\text{max}}},   \\\notag
\overline{\mathrm{C6}}&\!:
\mathop {\min }\limits_{
\bm  \Omega_{n}^{BE}} \sum\limits_{n = 1}^N {{\text{Tr}}[{\textbf{z}}_n^{\mathrm{H}}({{\dot {\textbf H}}}_{n}^{BT})^{\mathrm{H}}{ \bm {\mathcal{R}}}_\mathrm{n}^{ - 1}{{{{\dot {\textbf H}}}}_{n}^{BT}}{{\textbf{z}}_n}]}  \ge \frac{1}{{2{{\left| {{\beta _t}} \right|}^2}{\Gamma _{{\text{CRB}}}}}},\\\notag
{\mathrm{C8}}&\!:
{ {\textbf{W}} _{n,k}} \succeq \textbf{0},\forall n,k, \hspace{14mm} 
{\mathrm{C9}}:
{ {\textbf{V}_n}} \succeq \textbf{0},\forall n,\\\notag
{\mathrm{C10}}&\!:
\text{rank}({\textbf{W}_{n,k}}) \le 1,\forall n,k,\hspace{5mm} 
{\mathrm{C11}}:
\text{rank}(\textbf{V}_n) \le 1,\forall n,\\\notag
{\mathrm{C12}}&\!:\sum\limits_{n = 1}^N {{{\left\| {{\textbf{z}_n}} \right\|}^2} \le  } \sum\limits_{n = 1}^N ({\sum\limits_{k = 1}^K {\text{Tr}({{\overline {\textbf{W}} }_{n,k}})} }  + \text{Tr}(\textbf{V}_n)),
\end{align}
where the  variable  $\textbf{z}_n$ is introduced to facilitate the handling of constraint $\mathrm{C6}$.

In the sequel, a suboptimal scheme with low computational complexity is devised for problem optimisation in \eqref{recast_problem}.
To begin with, we first transform the constraint ${\mathrm{C1}}$ into its equivalent  formula
\begin{align}
	\hspace{-2mm}	{\mathrm{C1a}}:0 \le {s_{n,k}} \le 1,\forall n,k,
	{\mathrm{C1b}}:\sum\limits_{n = 1}^N {\sum\limits_{k = 1}^K {{s_{n,k}} - s_{n,k}^2 \le 0.} } 
\end{align}
However, a reverse convex function is introduced in ${\mathrm{C1b}}$.
To tackle this, 
we incorporate the constraint ${\mathrm{C1b}}$ into the objective function in the form of a penalty term as $\mathop {\min\limits_{k \in \cal K} {\overline{R}_k}} - \sum\limits_{n = 1}^N {\sum\limits_{k = 1}^K {\rho ({s_{n,k}} - s_{n,k}^2)} }, $
where the parameter ${\rho  \gg 1}$ serves as a penalty coefficient.

Next, note that the non-convexity inherent in the left-hand side (LHS) of constraints $ \overline{\mathrm{C4}}$ and $ \overline{\mathrm{C5}}$ poses a significant challenge to the resolution of \eqref{recast_problem}.
To tackle this, 
we resort to the Lagrangian dual transformation\cite{shen2018fractional},
so that we can convert the LHS in $ \overline{\mathrm{C4}}$ and $ \overline{\mathrm{C5}}$ into  more tractable forms as
\begin{align}
	\overline{\overline{\mathrm{C4}}}&:
	\label{c4}
	\mathop {\min }\limits_{\bm \Omega^{E}_{n,k} } \sum\limits_{n = 1}^N \big({{\log }_2}(1 + {\alpha _{n,k}}) - {\alpha _{n,k}} 	+ \frac{{(1 + {\alpha _{n,k}})\overline{\text{SINR}}_{n,k}}}{{1 + \overline{\text{SINR}}_{n,k}}}\big)  \notag \\
	&=		\mathop {\min }\limits_{\bm \Omega^{E}_{n,k} } 	\overline{\overline{R}}_k\ge {R_{{\text{min}}}}, \forall k, \\
	\overline{\overline{\mathrm{C5}}}&: \mathop {\max }\limits_{\bm  \Omega_{n}^{BE}} \sum\limits_{n = 1}^N \big({{\log }_2}(1 + {\gamma  _{n,k}}) - {\gamma  _{n,k}} + \frac{{(1 + {\gamma  _{n,k}})\overline{\text{SINR}}_{n,k}^{E} }}{{\overline{\text{SINR}}_{n,k}^{E} }}) \notag \\
	&\le R_{{\text{leakage}}}^i, \forall k, \forall i \label{c5}
\end{align}
where $\alpha _{n,k}$ and $\gamma _{n,k}$ refer to the introduced auxiliary variables.
Then, the problem \eqref{recast_problem} can be transformed as
\begin{align}&
	\hspace{-5mm}
	\label{easy_problem}
	\underset{
			{s_{n,k}, {\textbf{W}}_{n,k}, {{\overline {\textbf{W}} }_{n,k}},} {\textbf{V}_n, \alpha _{n,k}, \gamma _{n,k},\textbf{z}_n }
		}
		{\mathrm{maximize}}
\mathop {\min\limits_{k \in \cal K} {\overline{\overline{R}}_k}} - \sum\limits_{n = 1}^N {\sum\limits_{k = 1}^K {\rho ({s_{n,k}} - s_{n,k}^2)} }\\\notag
	\mathrm{s.t.} \, &\mathrm{C1a},\mathrm{C2},
	\overline{\mathrm{C3a}}, \overline{\mathrm{C3}},\overline{\overline{\mathrm{C4}}},  \overline{\overline{\mathrm{C5}}},
	\overline{\mathrm{C6}},
\mathrm{C7a-C7d},	{\mathrm{C8-C12}}.
\end{align} 

Note that the auxiliary variables  $\alpha _{n,k},\gamma _{n,k}$ are exclusively  present in constraints $\overline{\overline{\mathrm{C4}}}$ and $\overline{\overline{\mathrm{C5}}}$, respectively.
In light of this, we intend to iteratively optimize all the variables and derive the optimal solution of  the above auxiliary variables by leveraging the special structure of the problem.
Specifically, with given feasible points $ {{\overline {\textbf{W}} }^{j}_{n,k}}$ and $ \textbf{V}^{j}_n$  in the $j$-th iteration,
the optimal solutions for $\alpha _{n,k},$ and $ \gamma _{n,k}$ can be obtained by setting the derivatives of the LHS in $\overline{\overline{\mathrm{C4}}}$ and $\overline{\overline{\mathrm{C5}}}$ with respect to (w.r.t) $\alpha _{n,k}$ and $\gamma _{n,k}$, respectively, to zero. Thus we have
\begin{align}\label{afa}
	\alpha _{n,k}^ * = \overline{\text{SINR}}_{n,k},\forall n,k,\,\,
	\gamma _{n,k}^ * =\overline{\text{SINR}}_{n,k}^{E},\forall n,k.
\end{align}

Then, to handle the challenging issues caused by the fractional terms in \eqref{c4} and \eqref{c5}, we employ the quadratic transform to recast the constraints $\overline{\overline{\mathrm{C4}}}$ and $\overline{\overline{\mathrm{C5}}}$ into
\begin{align}\notag
	\widetilde {{\mathrm{C4}}}&:\mathop {\min }\limits_{\bm \Omega^{E}_{n,k} }\widetilde{R}_k=\mathop {\min }\limits_{\bm \Omega^{E}_{n,k} } \sum\limits_{n = 1}^N ( {\log _2}(1 + {\alpha _{n,k}}) - {\alpha _{n,k}} + 2{x_{n,k}}\\ \notag
	&\sqrt {(1 + {\alpha _{n,k}}){{ \eta }_{n,k}}}  - x_{n,k}^2\Big({\left| {{{({\textbf{h}}_{n,k}^{{E}})}^{{H}}}{{\textbf{x}}_{E}}} \right|^2} +\text{Tr}({\textbf{H}_{n,k}}{\textbf{V}_n})\\ &+ \sigma _{n,k}^2\Big)
	- \sum\limits_{{\left\| {{\textbf{h}_{n,j}}} \right\|^2} > {\left\| {{\textbf{h}_{n,k}}} \right\|^2}} {x_{n,j}^2{{ \eta }_{n,j}}} ) \ge {R_{{\text{min}}}},\forall k,\,\,\text{and}\\[-1.5mm]\notag
	\widetilde {{\mkern 1mu} {\mkern 1mu}\notag {\mathrm{C5}}}&:\mathop {\max }\limits_{\bm  \Omega_{n}^{BE}}  {\overline R}_k^{{E}}=\mathop {\max }\limits_{\bm  \Omega_{n}^{BE}} \sum\limits_{n = 1}^N ( {\log _2}(1 + {\gamma _{n,k}}) - {\gamma _{n,k}} + 2{z_{n,k}}\\\notag &
	\sqrt {(1 + {\gamma _{n,k}})\xi _{n,k}^n}  - z_{n,k}^2\xi _{n,k}^d)	
 \le R_{{\text{leakage}}}^i,\forall k,\forall i,
\end{align}
respectively,
where  ${ \eta }_{n,k}=\text{Tr}({\textbf{H}_{n,k}}{\overline {\textbf{W}} _{n,k}}), { \eta }_{n,j}=\text{Tr}({\textbf{H}_{n,k}}{\overline {\textbf{W}} _{n,j}}) $. Note that another two auxiliary variables, i.e.,  $x_{n,k}, z_{n,k}$, are introduced for the equivalence transformation.

In this case, the problem \eqref{recast_problem} is now equivalent to
\begin{align}& 
	\label{hei_problem}
 \hspace{-7mm} 
	\underset{s_{n,k}, {\textbf{W}}_{n,k}, {{\overline {\textbf{W}} }_{n,k}}, \textbf{V}_n,\textbf{z}_n,  x_{n,k}, z_{n,k} }{\mathrm{maximize}}\!\!
\mathop {\min\limits_{k \in \cal K} {\widetilde{R}_k}} - \sum\limits_{n = 1}^N {\sum\limits_{k = 1}^K {\rho ({s_{n,k}} - s_{n,k}^2)} }\\ \notag
	\mathrm{s.t.} \, &\mathrm{C1a},\mathrm{C2},
	\overline{\mathrm{C3a}}, 	\overline{\mathrm{C3}},\widetilde{\mathrm{C4}},  \widetilde{\mathrm{C5}},
	\overline{\mathrm{C6}},
\mathrm{C7a-C7d}, {\mathrm{C8-C12}}.
\end{align}
In what follows, after updating variables $\alpha _{n,k},$ and $ \gamma _{n,k}$ as in \eqref{afa},
we derive the optimal strategies for $ x_{n,k}$ and $ z_{n,k}$ with given feasible point
$ {{\overline {\textbf{W}} }_{n,k}}, \textbf{V}_n$ and $\alpha _{n,k}, \gamma _{n,k}$.
Specifically, as the derivatives of the LHS in $\widetilde{\mathrm{C4}}$ and $\widetilde{\mathrm{C5}}$ with respect to $ x_{n,k}$ and $ z_{n,k}$ are zero, respectively, the optimal $ x_{n,k}$ and $ z_{n,k}$ can be expressed as
\begin{align}\label{x}
	x_{n,k}^* &= \frac{{\sqrt {(1 + {\alpha _{n,k}}){{ \eta }_{n,k}}} }}{{\text{MUI}_{n,k} + {{\left| {{{({\textbf{h}}_{n,k}^{{E}})}^{{H}}}{{\textbf{x}}_{E}}} \right|}^2}+\text{Tr}({\textbf{H}_{n,k}}{\textbf{V}_n}) + \sigma _{n,k}^2}},\\
	z_{n,k}^* &= \frac{{\sqrt {(1 + {\gamma _{n,k}})\xi _{n,k}^n} }}{{\xi _{n,k}^d}},
	\label{z}\,\, \text{respectively}.
\end{align}

Now, given the optimal value for the above auxiliary variables, the problem \eqref{hei_problem} can be recast as
\begin{align}& 
	\hspace{-6mm} 
	\label{zhong_problem}
	\underset{s_{n,k}, {\textbf{W}}_{n,k}, {{\overline {\textbf{W}} }_{n,k}}, \textbf{V}_n,\textbf{z}_n}{\mathrm{maximize}}
\mathop {\min\limits_{k \in \cal K} {\widetilde{R}_k}} - \sum\limits_{n = 1}^N {\sum\limits_{k = 1}^K {\rho ({s_{n,k}} - s_{n,k}^2)} }\\ \notag
	\mathrm{s.t.} \, &\mathrm{C1a},\mathrm{C2},
	\overline{\mathrm{C3a}}, \overline{\mathrm{C3}},\widetilde{\mathrm{C4}},  \widetilde{\mathrm{C5}},
	\overline{\mathrm{C6}},
\mathrm{C7a-C7d}, {\mathrm{C8-C12}}.
\\[-7.5mm]\notag
\end{align}
Note that the constraint $\widetilde{\mathrm{C5}}$ and the objective function
are still not convex w.r.t. the optimised variables.
To handle this, we  adopt the iterative successive convex approximation (SCA) technique.
Specifically, for a given feasible solution ${{\overline {\textbf{W}} }^{j}_{n,k}}$ and $s^{j}_{n,k}$, the upper bound functions for  ${{\widetilde{R}}_{k}^{E}}$ and the penalty term can be obtained as
\begin{align}\label{SCA}\notag
	{\widetilde{R}}_{k}^{E} &\le {({\widetilde{R}}_{k}^{E})^{\text{ub}}}	= {({\overline R}_k^{{E}})^j}(\overline {\textbf{W}} _{n,k}^j) \\
&+ {\nabla _{{{\{ {{\overline {\textbf{W}} }_{n,k}}\} }_{r,c}}}}{\overline R}_k^{{E}} \times ({\{ {\overline {\textbf{W}} _{n,k}}\} _{r,c}} - {\{ \overline {\textbf{W}} _{n,k}^j\} _{r,c}}),
\end{align}
where 
\begin{align}
	{\nabla _{{{\{ {{\overline {\textbf{W}} }_{n,k}}\} }_{r,c}}}}\!\!\!{{\widetilde{R}}_{k}^{E}} \!=\!\! \sum\limits_{n = 1}^N \!\! \frac{{{z_{n,k}}\sqrt {(1 + {\gamma _{n,k}})} }}{{\sqrt {{{(\xi _{n,k}^n)}^j}} }}
	\!({\{ {{\textbf{H}}_{{{n}}}^{BE}}({\textbf{H}}_{{{n}}}^{BE})^{{H}}\} _{c,r}} ), 
	\\[-7.5mm]\notag
\end{align}
and
\begin{align}\notag
	{s_{n,k}} - s_{n,k}^2 &\le {(S_{n,k}^{\text{up}})^{{j}}}\\
	&= {s_{n,k}} - {(s_{n,k}^{{j}})^2} - 2s_{n,k}^{{j}}({s_{n,k}} - s_{n,k}^{{j}}),
\end{align}
respectively.

	Therefore, the problem \eqref{zhong_problem} can be updated with the following formula
\begin{align}
	& 
	\label{accept_problem}
	\underset{s_{n,k}, {\textbf{W}}_{n,k}, {{\overline {\textbf{W}} }_{n,k}}, \textbf{V}_n,\textbf{z}_n }{\mathrm{maximize}}\,\, 
\mathop {\min \limits_{k \in \cal K} {\widetilde{R}_k}}-\sum\limits_{n = 1}^N {\sum\limits_{k = 1}^K {\rho {(S_{n,k}^{\text{up}})^{{j}}}} } \\\notag
	\mathrm{s.t.} \, &\mathrm{C1a},\mathrm{C2},
	\overline{\mathrm{C3a}}, \overline{\mathrm{C3}}, \widetilde{\mathrm{C4}},  
	\overline{\mathrm{C6}},
	\mathrm{C7a-C7d},
	{\mathrm{C8-C12}},\\\notag
	\widehat {\mathrm{C5}}&:\mathop {\max }\limits_{\bm  \Omega_{n}^{BE}} {({\widetilde{R}}_{k}^{E})^{\text{ub}}} \le R_{{\text{leakage}}}^i, \forall k,\forall i.
\end{align}

In the following, note that the constraint $  \overline{\mathrm{C6}}$ is not convex w.r.t. the variable $\textbf{z}_n$. To tackle these,  we rewrite constraint  $  \overline{\mathrm{C6}}$  into the following expression and employ the penalty method  in the sequel 
	\begin{align}\notag
		\widetilde{\mathrm{C6}}&\!:\mathop {\min }\limits_{\bm  \Omega_{n}^{BE}} \sum\limits_{n = 1}^N(- {{\text{Tr}}[{\textbf{z}}_n^{\mathrm{H}}({{\dot {\textbf H}}}_{n}^{BT})^{\mathrm{H}}{ \bm {\mathcal{R}}}_\mathrm{n}^{ - 1}{{{{\dot {\textbf H}}}}_{n}^{BT}}{{\textbf{z}}_n}]})  + \frac{1}{{2{{\left| {{\beta _t}} \right|}^2}{\Gamma _{{\text{CRB}}}}}}\\
		&+{\chi _c}=0,	\hspace{30mm}\mathrm{C13}\!:{\chi _c} \ge 0,
	\end{align}
	where ${\chi _c}$  is the introduced slack variable. Afterwards, via  incorporating the constraint  $\widetilde{\mathrm{C6}}$  into the objective function as penalty term, we derive the  subsequent expression $		\mathcal{O}=\mathop {\min \limits_{k \in \cal K} {\widetilde{R}_k}}-\sum\limits_{n = 1}^N {\sum\limits_{k = 1}^K {\rho {(S_{n,k}^{\text{up}})^{{j}}}} } -\rho_c\Big(\mathop {\min }\limits_{{\bm  \Omega_{n}^{BE}}} \sum\limits_{n = 1}^N(- {{\text{Tr}}[{\textbf{z}}_n^{\mathrm{H}}({{\dot {\textbf H}}}_{n}^{BT})^{\mathrm{H}}{ \bm {\mathcal{R}}}_\mathrm{n}^{ - 1}{{{{\dot {\textbf H}}}}_{n}^{BT}}{{\textbf{z}}_n}]}) + \frac{1}{{2{{\left| {{\beta _t}} \right|}^2}{\Gamma _{{\text{CRB}}}}}}+{\chi _c}\Big)^2,$
	where  $\rho_c\gg 1$ denotes the  penalty factor.
	
	Besides, for a given feasible solution $\{{{\overline {\textbf{W}} }_{n,k}},\textbf{V}_n,\textbf{z}_n\}$ in each iteration, ${\chi _c} $ can be updated as $	{\chi _c}=\mathop {{\max}}\{  \mathop {\min }\limits_{{\bm  \Omega_{n}^{BE} }} \sum\limits_{n = 1}^N {{\text{Tr}}[{\textbf{z}}_n^{\mathrm{H}}({{\dot {\textbf H}}}_{n}^{BT})^{\mathrm{H}}{ \bm {\mathcal{R}}}_\mathrm{n}^{ - 1}{{{{\dot {\textbf H}}}}_{n}^{BT}}{{\textbf{z}}_n}]}-\frac{1}{{2{{\left| {{\beta _t}} \right|}^2}{\Gamma _{{\text{CRB}}}}}},0\}.$

So far,  the optimised  problem is convex w.r.t the optimization variables except for the rank-one constraints ${\mathrm{C10}}$ and  ${\mathrm{C11}}$.
To circumvent this issue, the semidefinite relaxation (SDR) technique\cite{hu2021robust,hu2020sum} is utilized via dropping them from the formulated  problem.
Thus, we yield
\begin{align}
	& 
	\label{accept3_problem}
	\underset{s_{n,k}, {\textbf{W}}_{n,k}, {{\overline {\textbf{W}} }_{n,k}}, \textbf{V}_n,\textbf{z}_n }{\mathrm{maximize}}\,\, 
\mathcal{O}\\\notag
	\mathrm{s.t.} \, &\mathrm{C1a},\mathrm{C2},
	\overline{\mathrm{C3}}, \widetilde{\mathrm{C4}},  \widehat {\mathrm{C5}},
\mathrm{C7a-C7d},
	{\mathrm{C8-C9}},{\mathrm{C12}},
\end{align}
where  the problem can be efficiently solved by CVX.
To narrow the gap between the upper bound and the optimal objective function, we iteratively  update the feasible point
$\{s_{n,k}^{{j}}, {\overline {\textbf{W}}^{j} _{n,k}},\textbf{V}_n^j\}$ via solving the problem \eqref{accept3_problem} iteratively  until it satisfies the convergence condition.

\section{Simulation Results}
This section provides the numerical results of our proposed scheme (PS) under different setups. 
Throughout the simulations,  it is assumed that $K=5$, $N=2$, $V=18$,
$P_{BS}^{{\text{max}}} = 30{{ }}$ dBm, $M=2$, $P_{E}=10$ dBm, $B=0.5$ MHz, $K_N=3$, $\sigma _{n,k}^2 = \sigma _{E}^2 =  - 110{{ }}$ dBm, $R_\text{min}=1.6$ bit/s/Hz
and ${\Gamma _{{\text{CRB}}}}=0.01$. 
We consider the Rayleigh small-scale fading with unit variance and the large-scale fading is $ - 128.1 - 37.6{\log _{10}}({d}[{\text{km}}])$ dB. 
There are $L=3$ SLs in the system and the corresponding maximum tolerable leakage rates are $R_{{\text{leakage}}}^1=0.5 $ bit/s/Hz,  $R_{{\text{leakage}}}^2=1 $ bit/s/Hz, and  $R_{{\text{leakage}}}^3=1.5 $ bit/s/Hz, respectively.
We assume that the SL of users 1 and 2 is SL3, the SL of users 3 and 4 is SL2, and that of user 5 is SL1.
 The directions and the distances from the BS to the users and the target are randomly distributed at (0°, 180°) and (240, 390) m, respectively.
 We assume the potential direction of the active Eve is 0° and its distance to the BS is randomly selected within the range of (360, 380) m.
 For ease of notation, we introduce CSI error parameters 
 ${\chi ^{BE}_{{{n}}}} = \frac{{{\varepsilon^{BE} _{{{n}}}}}}{{{{\left\| {{\textbf{H}^{BE}_{{{n}}}}} \right\|}_F}}}$ and $\chi _{{{n,k}}}^{{E}} = \frac{{\varepsilon _{{{n,k}}}^{{E}}}}{{{{\left\| {\textbf{h}_{{{n,k}}}^{{E}}} \right\|}_2}}}$ to denote the maximum normalized estimation error of channels ${\textbf{H}^{BE}_{{{n}}}}$and $\textbf{h}_{{{n,k}}}^{{E}}$, respectively.
 Unless otherwise stated, we set ${\chi^{BE} _{{{n}}}} = \chi _{{{n,k}}}^{{E}}  = \chi =0.02$.
For comparison, our proposed scheme will be compared with the following baselines:
1) \textit{\textbf{No AN}}: where the AN technique is not exploited; 
2) \textit{\textbf{NS AN}}: where the AN is produced in the null space of the users’ channels;
3) \textit{\textbf{SC-NOMA}}: where a single-subcarrier NOMA assisted ISAC system is considered.

\begin{figure}
	\centering
	\begin{minipage}{0.22 \textwidth}
		\centering
		\includegraphics[width=1.7 in]{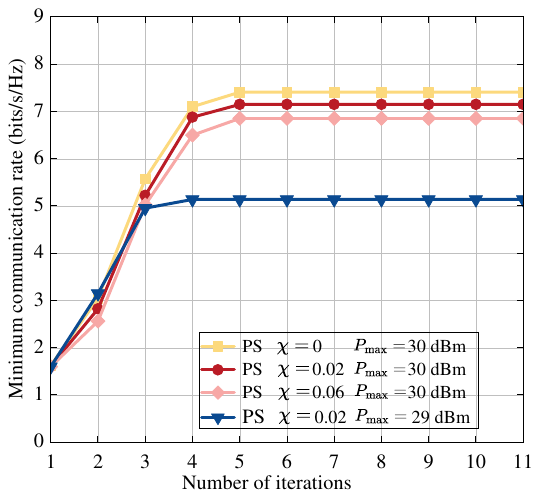}
		\caption{The convergence of the proposed scheme under different settings.}
		\label{convergence}
	\end{minipage}
	\quad
	\begin{minipage}{0.22 \textwidth}
		\centering
		\vspace{-0.32cm} 
		\includegraphics[width=1.7 in]{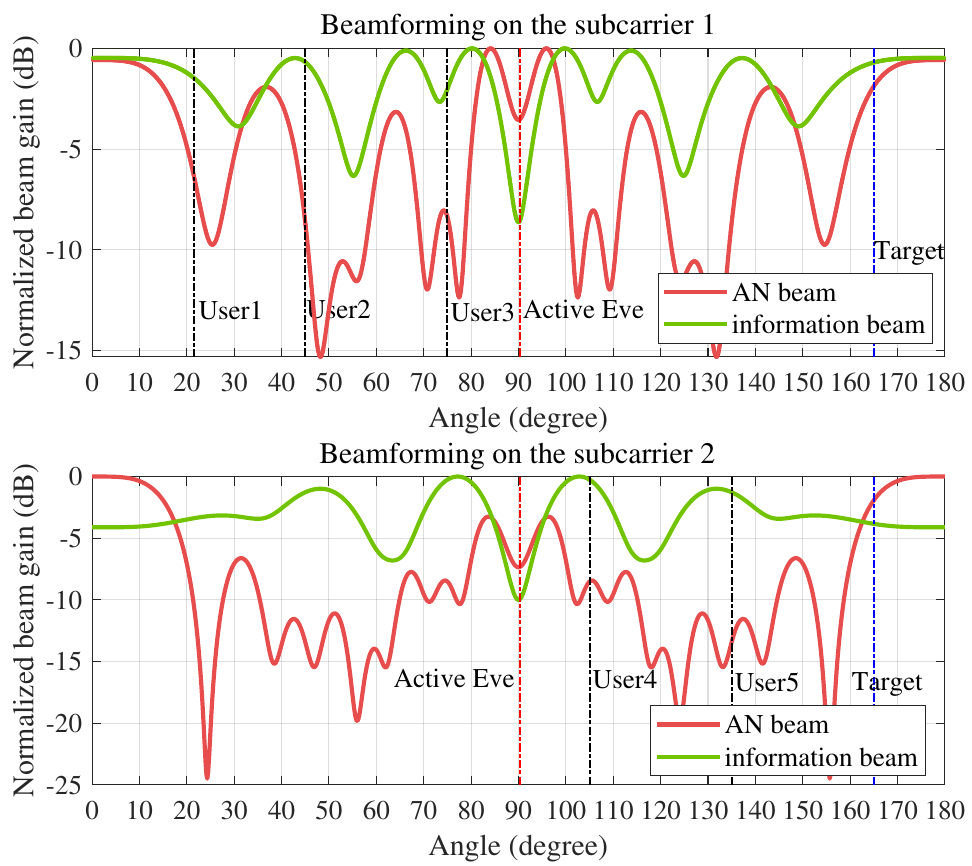}
		\caption{{The proposed secure beamforming versus angle.}}
		\label{beamforming}
	\end{minipage}
\end{figure}	
\begin{figure}
	\centering
	\begin{minipage}{0.22 \textwidth}
		\centering
		\includegraphics[width=1.7in]{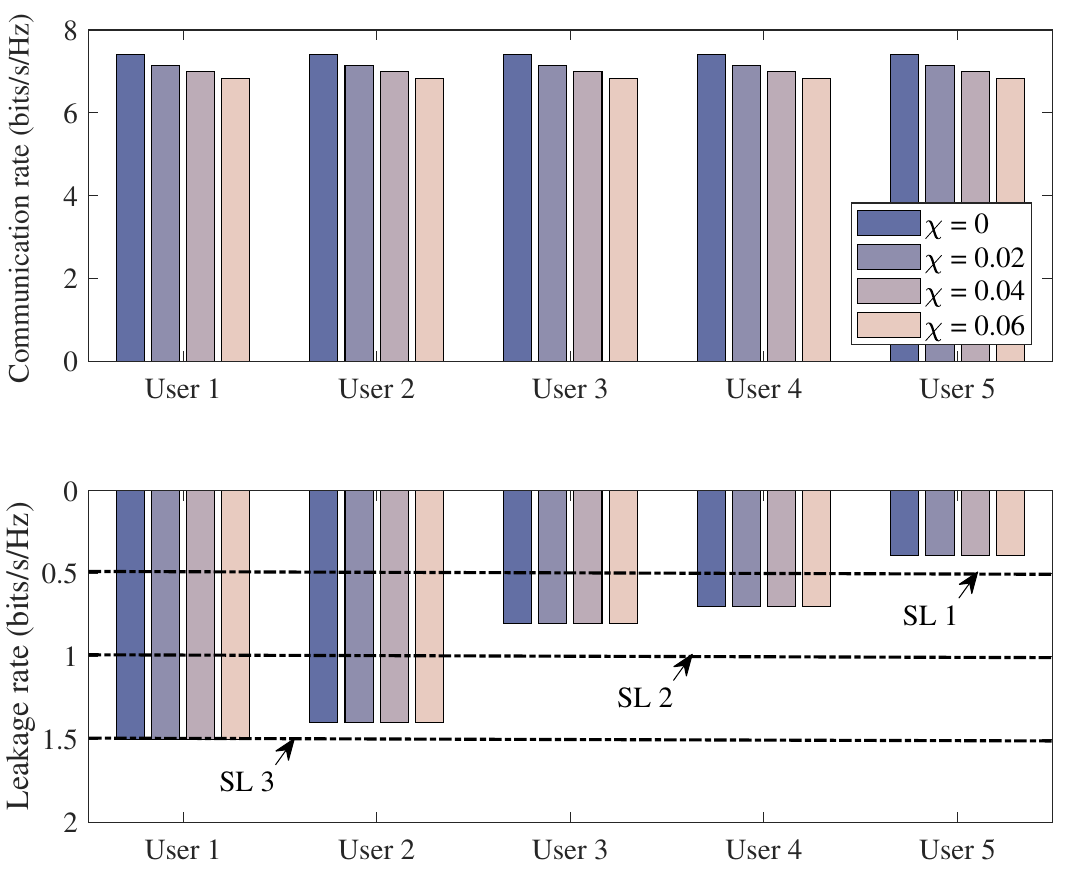}
		\caption{The communication and leakage rates for each user under different CSI errors.}
		\label{Leakage}
	\end{minipage}
	\quad
	\begin{minipage}{0.22\textwidth}
		\centering
		\vspace{0.15cm} 
	\setlength{\abovecaptionskip}{0.15cm} 
		\includegraphics[width=1.7 in]{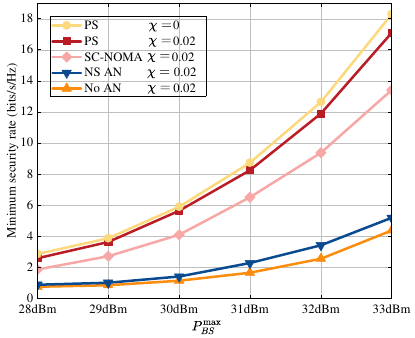}
		\caption{The minimum security rate versus $P_{BS}^{{\text{max}}}$ for the PS and benchmarks.}
		\label{Performance}
	\end{minipage}
\end{figure}

Fig. \ref*{convergence} depicts the convergence of the proposed scheme under different settings.
As clearly shown, the PS achieves convergence within six iterations under various system settings,  ensuring the proposed algorithm’s convergence. Additionally, it is evident that increasing the transmit power of BS and improving channel certainty are beneficial for system performance enhancement.

Fig. \ref*{beamforming}  illustrates the secure beamforming of our proposed scheme.
It can be observed that the information beam is stronger than the AN beam at all communication users’ directions, ensuring better SNR and maximizing users' rates. 
Meanwhile, in the vicinity of the target, both the information and AN beams reveal high normalized amplitudes to meet the position estimation requirements. 
More remarkably, at the potential active Eve's direction,  the information beam reaches its local minimum and the AN beam is noticeably stronger,  suppressing eavesdropping on the confidential information.
This verifies the effectiveness of our secure beamforming design.

Fig. \ref*{Leakage} presents the communication and leakage rates for each user under different CSI errors.
We identify that all the users achieve the same communication rate across varying channel conditions, which ensures user fairness.
Notably, thanks to the beamforming design, 
the leakage rate for each user remains below its corresponding leakage threshold as the CSI error increases,  fulfilling graded confidentiality requirements and validating our proposed strategy.
Additionally, the stability of leakage rates across varying channel uncertainties highlights the system’s security robustness.
Besides, the communication rates decrease with increasing channel uncertainty, since the system needs to expend more power to neutralize the impact of higher CSI uncertainty.

In Fig. \ref*{Performance}, we present a comparison of the minimum security rate between the PS and other baselines under different values of $P_{BS}^{{\text{max}}}$.
It is evident from the figure that the security rates of all schemes increase gradually with the increased value of $P_{BS}^{{\text{max}}}$. Additionally, with a given transmit power, the  PS outperforms other baselines significantly.  Specifically, compared to the No AN scheme, the PS  shows superior security performance due to the introduction of AN, which effectively reduces the quality of Eve’s eavesdropping channel. 
More importantly, the comparable security performance of the No AN and the NS-AN schemes highlights  the significance of AN optimization. 
By optimizing the  AN's beamforming, the SNR gap between the active Eve  and the legitimate user is maximized, fully leveraging the benefits of AN technology and enhancing security.
Furthermore, the PS demonstrates superior security performance compared to the SC-NOMA scheme.
The reason is that the subcarrier allocation in PS provides higher design flexibility and enables more effective use of the available bandwidth,  enhancing the system's performance.

\section{Conclusion}
In this paper, we explored the secure beamforming and subcarrier allocation in an MC-NOMA-empowered ISAC system against an active Eve under imperfect CSI.
We aimed to maximize the minimum   communication rate to ensure user-fairness, while satisfying sensing performance and graded security demands.
An efficient  SCA-based algorithm was proposed to solve the challenging problem.
Simulation results demonstrate that the PS enhances the reliability of the considered system and provides high robustness.

\end{document}